\documentclass[12pt]{iopart}
\usepackage{iopams}
\usepackage{graphicx}
\usepackage{caption}
\usepackage{subcaption}
\usepackage{hyperref}
\usepackage{hypernat}
\usepackage{color}
\usepackage{lineno} 
\usepackage{cite}

\begin{document}
\title[Modeling urban mobility in Bogotá’s public transport system]{From motifs to Lévy flights: Modeling urban mobility in Bogotá’s public transport system}
\author{Juan F. Alayón-Martínez${}^1$, Alejandro P. Riascos$^{1}$}

\address{${}^{1}$ Departamento de F\'isica, Universidad Nacional de Colombia, Bogotá, Colombia}

\begin{abstract} In this paper, we study two years of access card validation records from Bogotá’s multimodal public transport system, comprising over 2.3 billion trips across bus rapid transit, feeder buses, dual-service buses, and an aerial cable network. From user trajectories constructed exclusively from access records, we derive motifs that reveal recurrent mobility patterns extending beyond simple two-location visits. This approach enables the construction of an integrated origin--destination (OD) matrix covering 2,828 urban zones. Similarity analysis using the Jensen--Shannon divergence confirms the temporal stability of mobility structures across semesters, despite infrastructure changes and fare policy adjustments. From the obtained OD matrices, we derive transition probabilities between zones and uncover a robust power-law relationship with geographical distance, consistent with Lévy flight dynamics. We validate our model using Monte Carlo simulations showing that reproduces both local and long-range displacements, with similar scaling exponents across time. These findings demonstrate that Bogotá's public transport mobility can be effectively modeled through Lévy processes, providing a novel framework for analyzing complex transportation systems based solely on user access records.
\end{abstract}

\maketitle

\section{Introduction}
The study of human mobility in cities is both important and challenging, as more than half of the world’s population now resides in urban areas \cite{BattyBook_2013}. Advances in mobile and digital platforms have made it possible to examine human mobility in detail through the digital traces they generate \cite{barthelemy_2016,Barbosa2018}. Identifying global mobility patterns is crucial for applications in urban planning, transportation systems, and the analysis of how a city’s spatial distribution shapes mobility \cite{OrtuWill2011,Louail_SciRep2014,Louail_NatCommun2015,Lee2015}, as well as for understanding the encounter and contact networks that emerge \cite{RiascosMateosPlos2017}.
\\[2mm]
Furthermore, the recent availability of data records for diverse aspects of daily urban life has enabled the detailed characterization of human movement, allowing the identification of behavioral patterns at different scales \cite{Gonzalez2008,Alessandretti2018,Alessandretti2020,Martinez-Gonzalez2022,Pappalardo2023,Betancourt2023,Martinez_BRT_2024}. In the context of urban mobility and public transportation systems, smart-card automated fare collection records provide consistent, longitudinal evidence on ridership and allow analysts to characterize demand, service performance, and user behavior at spatial and temporal scales \cite{Pelletier2011}. Complementary data streams, such as mobile-phone call detail records have been used to infer origin–destination (OD) flows, trip purposes, and diurnal patterns at the population scale, revealing both regularities and heterogeneity in individual trajectories \cite{Calabrese2013,Iqbal2014,Alexander2015,yabe2025behaviour}. In this manner, big data in urban transportation systems supports the analysis of passenger behavior, operation optimization, and policy applications, enabling demand forecasting, timetable rescheduling, network planning, and fare policy design for efficient and sustainable mobility \cite{Lu2021}.
\\[2mm]
Origin–destination matrices underpin mobility analysis \cite{Hussain2021,Lei2021}. Smart-card records make it possible to estimate OD matrices by processing the raw validation data, identifying when passengers enter and leave the system, detecting transfers between services, grouping stops into larger spatial units, and ensuring the overall consistency of the reconstructed trips \cite{Hussain2021}. Persistent challenges include assumptions related to trip chaining, sensitivity to threshold selection, boundary effects, biases, scalability to large populations, and limited validation \cite{Hussain2021}. Methodological innovations include a minimum entropy rate framework that infers alighting stops for single trips while preserving passengers’ travel regularity, reducing noise and improving OD completeness for subsequent mobility analyses \cite{Lei2021}. Another contribution is an enhanced trip-chain method that addresses single and unlinked trips, calibrates transfer thresholds, aggregates stop-to-zone flows, and applicability across both bus and subway networks \cite{Radfar2025}. In addition, predictive modeling of OD dynamics increasingly exploits spatiotemporal learning to capture network effects, nonlinearity, and regime changes. Graph-convolutional formulations treat OD matrices as signals on networks, enabling the propagation of spatial context and improving short- and medium-horizon forecasts \cite{Wang2019}. Recent architectures further refine local and global dependencies with fine-grained multilayer perceptrons tailored to OD tensors, improving accuracy while controlling model complexity \cite{Liu2025}. In urban rail and metro settings, short-term OD passenger-flow prediction frameworks demonstrate practical utility for operations and crowd management \cite{Su2024}. Taken together, these developments highlight the need for OD-estimation methods that remain reliable across different contexts by relying on transparent modeling assumptions, incorporating information on passenger behavior, and validating results using independent data sources whenever available \cite{MOHAMMED2023315}.
\\[2mm]
On the other hand, several studies have shown that human mobility exhibits long-range dynamics similar to L\'evy walks, a strategy also observed in many animal species and in humans \cite{Barbosa2018,BoyerPRS2006}. In network contexts, L\'evy flights demonstrate that long-range displacements enhance the ability to efficiently reach any site by inducing the small-world property through the dynamics \cite{RiascosMateosLF2012}. This mechanism has subsequently been investigated in diverse settings, including fractional diffusive transport \cite{FractionalBook2019}, dynamics on multiplex networks \cite{Guo2016}, human mobility \cite{RiascosMateosPlos2017,LoaizaPlosOne2019,Neira2024_SciRep}, and semi-supervised learning \cite{SdeNigris2017}, among others \cite{FractionalBook2019,reviewjcn_2021}.
\\[2mm]
In this paper, we present a comprehensive study of urban mobility patterns within Bogotá's public transportation system by leveraging a rich dataset consisting exclusively of access card records collected over two years. We begin by characterizing daily and spatial usage patterns, identifying stable and recurrent motifs of user trajectories that reflect the complex nature of the system. Building on these insights, we construct a detailed OD matrix that integrates multiple trip steps, overcoming limitations of prior approaches that considered only simple two-step trips. We then analyze the relationship between transition probabilities and geographic distances of users trips, revealing a robust scaling behavior that motivates the modeling of user displacements as Lévy flights on a network of urban zones. Then, by using Monte Carlo simulations, we validate that the Lévy flight dynamics successfully reproduce the empirical distribution of interzonal travel distances, capturing both local and long-range mobility dynamics. Our findings demonstrate the stability of mobility patterns despite temporal changes in the transport infrastructure and provide a novel quantitative framework for understanding complex urban movements. This research offers diverse tools for the analysis of mobility patterns in transportation systems using only access validation records.
\section{Methods}
\subsection{Dataset description}
\begin{table}[b!]
		\centering
		\caption{ \label{Tabla_1}
			{\bf Percentage of validation access card uses in the public transport system of Bogotá.} The dataset covers two years of data collection, divided into four semesters (Semester I: July–December 2023, Semester II: January–June 2024, Semester III: July–December 2024, and Semester IV: January–June 2025). The table reports the percentage distribution of users according to the number of daily trips, ranging from one to seven or more. It also includes the total number of trips per semester and the percentage of trips made through transfers, defined as journeys that allow passengers, with a single fare, to take up to two feeder buses and one BRT service within 125 minutes at the same price. Percentages correspond to the relative participation of each category within each semester.}
		    \resizebox{\textwidth}{!}{
			\begin{tabular}{|c|c|c|c|c|c|c|c|c|}
				\hline
				\multicolumn{1}{|c|}{\bf Daily use} & \multicolumn{2}{c|}{\bf Semester I} & \multicolumn{2}{c|}{\bf Semester II} & \multicolumn{2}{c|}{\bf Semester III} & \multicolumn{2}{c|}{\bf Semester IV} \\ \hline
				\multicolumn{1}{|c|}{} & \multicolumn{1}{c|}{\bf Quantity} & \multicolumn{1}{c|}{\bf \%} & 
				\multicolumn{1}{c|}{\bf Quantity} & \multicolumn{1}{c|}{\bf \%} &
				\multicolumn{1}{c|}{\bf Quantity} & \multicolumn{1}{c|}{\bf \%} &
				\multicolumn{1}{c|}{\bf Quantity} & \multicolumn{1}{c|}{\bf \%} \\ \hline
				1 Use        & 79,318,543  & 13.53 & 72,107,310  & 13.30 & 77,564,163  & 12.71 & 72,259,030  & 12.76 \\ \hline
				2 Uses       & 216,600,274 & 36.94 & 201,854,542 & 37.23 & 224,987,774 & 36.86 & 210,031,776 & 37.08 \\ \hline
				3 Uses       & 121,315,356 & 20.69 & 110,462,991 & 20.37 & 124,773,723 & 20.44 & 115,466,970 & 20.39 \\ \hline
				4 Uses       & 103,600,240 & 17.67 & 97,647,848  & 18.01 & 111,845,280 & 18.32 & 104,336,836 & 18.42 \\ \hline
				5 Uses       & 36,830,805  & 6.28  & 34,074,785  & 6.28  & 39,891,830  & 6.54  & 37,082,375  & 6.55 \\ \hline
				6 Uses       & 17,636,352  & 3.01  & 16,169,340  & 2.98  & 19,337,514  & 3.17  & 17,188,422  & 3.03 \\ \hline
				7 Uses or more & 11,124,068  & 1.90  & 9,871,041   & 1.82  & 11,979,044  & 1.96  & 10,021,929  & 1.77 \\ \hline
				\bf Total    & \multicolumn{2}{c|}{586,425,638} & \multicolumn{2}{c|}{542,187,857} & \multicolumn{2}{c|}{610,379,328} & \multicolumn{2}{c|}{566,387,338} \\ \hline
				Transfers   & 109,056,446 & 18.60\% & 103,323,821 & 19.06\% & 117,075,238 & 19.18\% & 113,875,148 & 20.11\% \\ \hline
			\end{tabular}
		}
\end{table}
\begin{figure}[!t]
	\centering
	\includegraphics[width=1\linewidth]{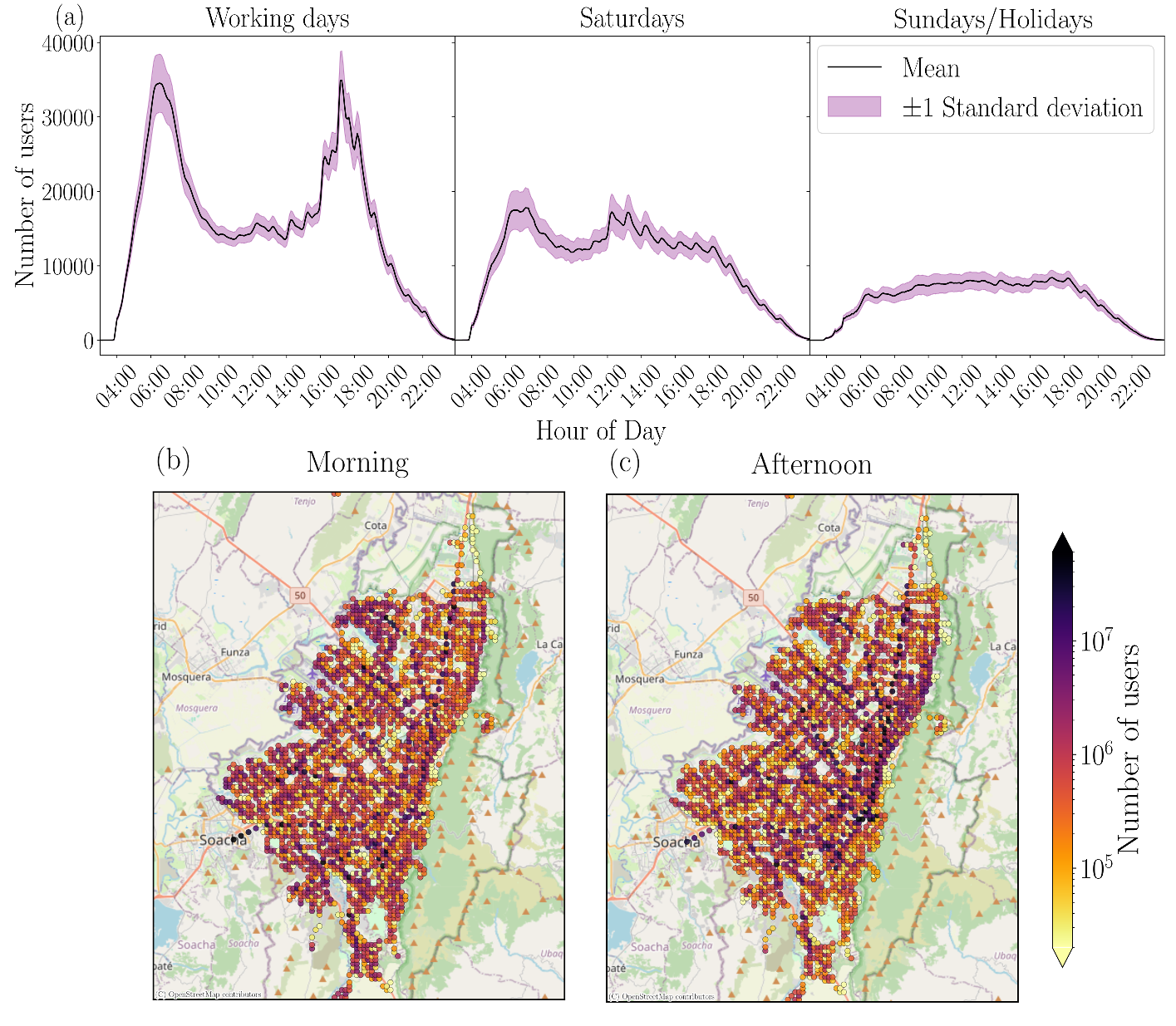}
	\caption{{\bf 
			Analysis of mobility patterns of public transport users in Bogotá.} (a) Frequency of use throughout the day in five-minute intervals. The black line represents the average, and the shaded region indicates the standard deviation of the data for weekdays, Saturdays, and  Sundays/Holidays, highlighting the strong regularity of the city, particularly on weekdays. (b) Heatmap showing the spatial distribution of trips during morning hours and (c) afternoon hours. Panels (b)-(c) show the city divided into 2,828 zones of 300 × 300 meters, each containing bus stops and stations across transport modes. The color bar, presented on a logarithmic scale, indicates the number of users, highlighting the urban structure and the predominance of the BRT system over feeder buses.
	}
	\label{Fig_1}
\end{figure}
Understanding human mobility within the SITP (Integrated Public Transport System, by its initials in Spanish) in Bogotá is particularly relevant because the city operates without a complementary metro network. The general operating hours of the SITP are from Monday to Saturday between 4:00 a.m. and 11:00 p.m., and on Sundays and holidays between 5:00 a.m. and 10:00 p.m., although these hours may vary across specific routes. The system serves approximately eight million inhabitants, reaching nearly ten million when surrounding municipalities are included. Bogotá’s public transport comprises four modes: a bus rapid transit (BRT) network, known as TransMilenio, with 148 stations and 114 km of exclusive lanes; feeder buses (SITP zonal) serving more than 7,500 stops; dual-service buses operating in both exclusive and mixed lanes; and an aerial cable system with three stations.  Access is provided through a smart card system that also enables transfers between modes: passengers may use one BRT service and up to two feeder buses within a 125-minute window by paying a single fare. All data are publicly available with access granted by TransMilenio  \cite{Datos_Transmilenio}.
\\[2mm]
The data used in this study come from validation records of access cards, which provide precise time and location information for every entry into the system. These records were collected between July 2023 and June 2025 and grouped into four semesters: Semester I (July–December 2023), Semester II (January–June 2024), Semester III (July–December 2024), and Semester IV (January–June 2025). Each validation event contains the entry location, date and time, fare paid, and a fully anonymized yet persistent unique identifier. Importantly, this identifier remains stable over the entire period of observation, ensuring that validations can be consistently associated with the same physical card and enabling reliable trajectory reconstruction. 
\\[2mm]
Table \ref{Tabla_1} presents a general summary of the data per semester, including the daily frequency of  the most relevant card use 
(1 use to 7 uses or more), which reveals a marked trend toward two trips per day (37.02\% of the total records). However, between three and six trips concentrate a considerable share of 48.04\%. This pattern can be partly explained by the transfer system and by specific mobility needs, which introduce greater diversity in travel behaviors. In addition, approximately one-fifth of the records correspond to transfers. In total, this study analyzes 2,305,380,161 records.
\\[2mm]
To further analyze the data, we first examine how users move across the city in order to better understand its dynamics. The results are presented in Fig. \ref{Fig_1}. Panel \ref{Fig_1}(a) shows the distribution of card validations throughout the day, obtained from the exact validation times, for three categories: weekdays, Saturdays, and Sundays/Holidays. The shaded region  corresponds to the standard deviation, while the black line shows the average. The results show that weekdays exhibit two pronounced peaks (around 6:00 a.m. and 5:30 p.m.). Saturdays, however, display a distinct pattern that differs from those reported for cities such as New York or Chicago \cite{LoaizaPlosOne2019}, as Bogotá exhibits an intermediate behavior between a working day and a day of rest. Sundays and holidays were grouped together, as they display similar dynamics, with fewer passengers. This behavior is consistent with previous studies of daily activity in South American cities \cite{Betancourt2023}, indicating that Bogotá follows highly regular patterns from Monday to Friday throughout the year. Thus, a stable mobility pattern can be identified for weekdays. Based on this observation, weekdays provide the most representative and robust basis for the subsequent analysis of Bogotá’s urban dynamics, as they are more homogeneous and account for the largest fraction of records, corresponding to 80.74\% of the total over the two-year period.
\\[2mm]
In addition, the geographical area of the city was partitioned into square zones of 300 m × 300 m to capture the most relevant mobility patterns and to aggregate multiple transport stations within the same area. This procedure reduced more than 7,600 individual stops to 2,828 zones with nonzero records in the dataset. Figure \ref{Fig_1}(b) displays the spatial distribution of validations during morning hours, while Fig. \ref{Fig_1}(c) shows the corresponding distribution in the afternoon, after 12:00 p.m. In both cases, the color bar represents frequencies on a logarithmic scale. The areas with the highest intensity correspond primarily to the BRT system, effectively outlining its network. In the morning, concentrations occur in peripheral areas near terminals and connection stations, whereas in the afternoon they shift toward commercial and work areas in the city center. This pattern also appears, although less prominently, in areas without BRT infrastructure. These findings reveal a marked spatial organization of the city, with peripheral areas concentrating residential locations and the center concentrating work, educational, and commercial activities. This suggests that daily trips are strongly oriented toward these zones, regardless of the number of system uses per day.
\subsection{Mobility analysis through subgraph motifs}

\begin{figure}[!t]
	\centering
	\includegraphics[width=1\linewidth]{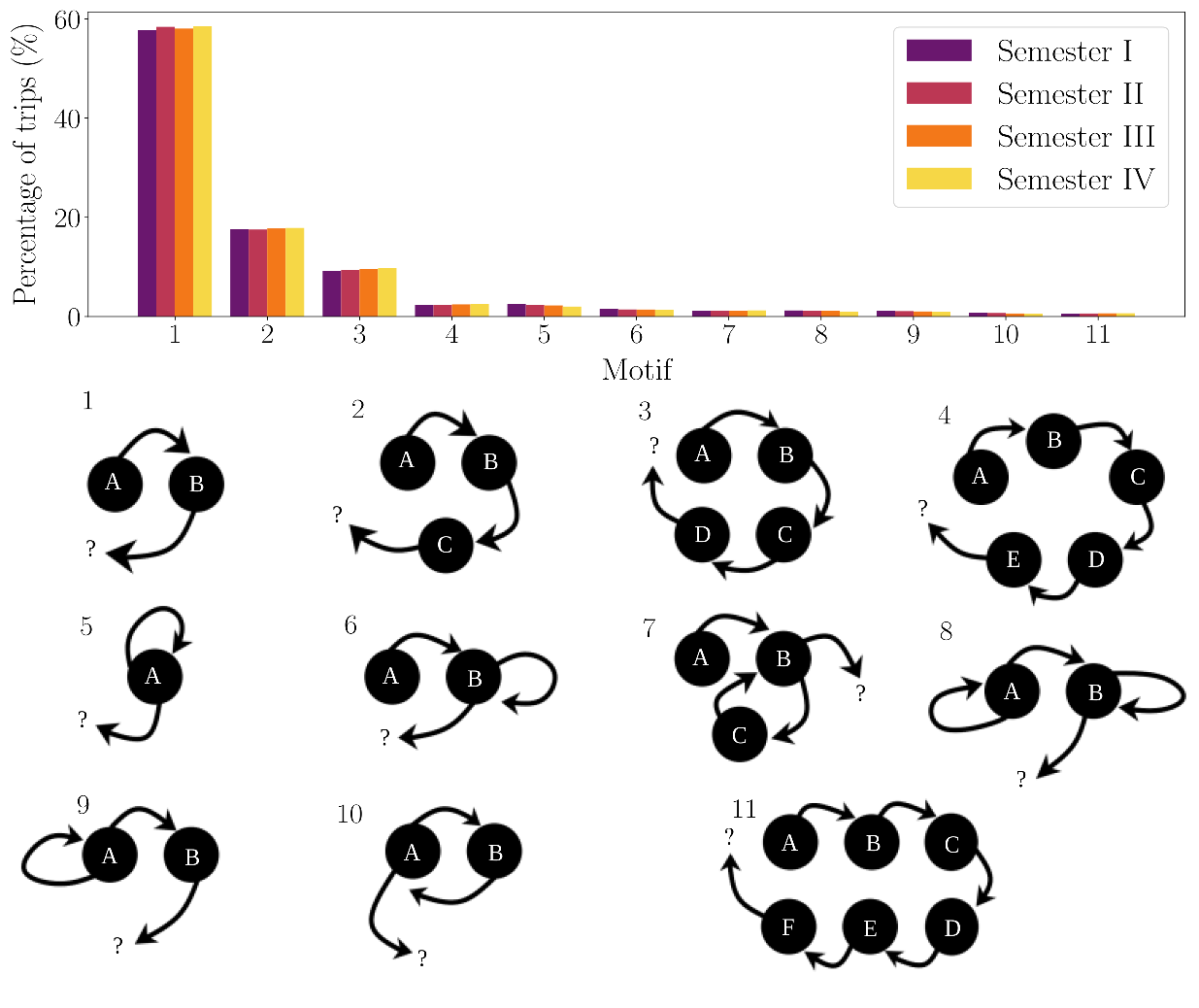}
	\caption{{\bf 
			Statistical representation of the 11 most frequent motifs among system users.} These motifs account for 96.06\% of all reconstructed trajectories, which correspond to chains of two to six trips with successive accesses at locations denoted $A, B, \ldots, F$. Each number identifies the diagram of its respective motif, constructed under the assumption that each validation represents the immediate destination of the previous trip; therefore, the final destination remains undefined. The most recurrent pattern corresponds to linear trajectories across different areas without repetitions, highlighting the need to further investigate mobility structures more complex than the sequence of trips $A\to B\to A$. Motifs represent observed intermediate sequences of validations and do not provide information about the final destination of the trajectory.}
	\label{Fig_2}
\end{figure}

The dataset of access validations using unique codes makes it possible to identify the specific trips made by each user. These trips can be represented as subgraphs and, following the adopted definition, are referred to as motifs. Motifs capture recurrent patterns in individual mobility \cite{gonzalez} and enable the classification of user behavior within the transportation system. Following this approach, the daily trips of each user are represented as a single motif that aggregates all accesses made during that day. The complete collection of motifs obtained in each semester was then grouped and classified to provide a clearer characterization of user displacements, as shown in Fig. \ref{Fig_2}. For each semester, we calculated the percentage associated with the 11 most frequent motifs of urban trajectories identified in our analysis. It is important to note that, since the Bogotá system requires a single validation only at the point of entry, the construction of motifs requires at least two consecutive validations, where the destination of one trip is considered the origin of the next. Consequently, motifs do not have a fixed and explicitly identifiable destination, as there is no reliable criterion to determine the precise endpoint of each trip.
Figure \ref{Fig_2} shows the bar distribution of each motif, identified by a number. The eleven main motifs represent 96.06$\%$ of all reconstructible trajectories in the city of Bogotá involving two or more access card validations. A clear tendency toward trajectories with between two and six validations is observed, with two- and three-step trips being the most frequent, consistent with the results reported in Table \ref{Tabla_1}. Moreover, there is a remarkable homogeneity across semesters, indicating that mobility patterns exhibit temporal stability rather than significant seasonal changes, thereby reinforcing the hypothesis that mobility in Bogotá is shaped by a persistent urban structure. Regarding two-location motifs, these account for nearly 58$\%$ of all trajectories, underscoring the importance of considering other types of trips as well. Excluding them would imply a substantial loss of information, since users exhibit differentiated mobility behaviors. Finally, for the subsequent stages of the analysis, a data-cleaning process was performed to remove corrupted records associated with non-representative transactions, such as illegal ticket sales.

\subsection{Characterizing mobility through the Origin–Destination matrix}
\label{Sec_ODmatrix}
\begin{figure}[!t]
	\centering
	\includegraphics[width=1\linewidth]{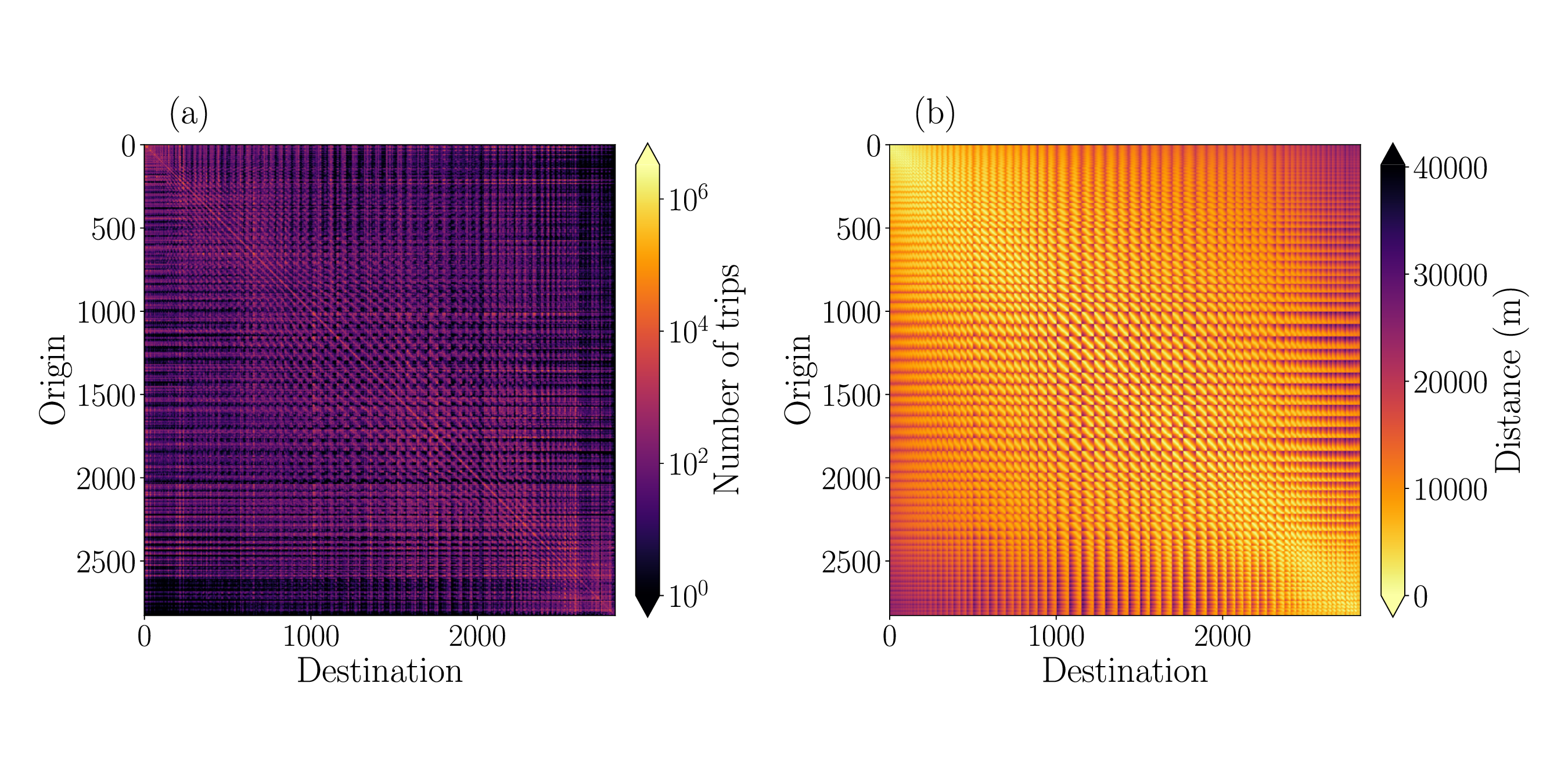}
	\caption{{\bf Matrix representation of public transport mobility in Bogotá.}  
		(a) OD matrix representing the movement of public transport users across 2,828 zones, corresponding to the 300 m × 300 m partition shown in Fig. \ref{Fig_1}. The matrix includes trajectories of two to six trips, which account for 85.04\% of the data. The OD matrix was constructed following the motif logic, where the destination corresponds to the immediate subsequent trip.  (b) Distance matrix of Bogotá, where values, expressed in meters, represent the geographical distance between origin and destination zones, as indicated by the color bar.}
	\label{Fig_3}
\end{figure}

An abstract representation of human movements and their interactions with places or objects is given by OD matrices, which are widely used to uncover interaction patterns between people and their territories, as well as in urban planning processes. These matrices have been constructed using various approaches: from data sources that inherently contain explicit origins and destinations \cite{Riascos_taxis,LoaizaPlosOne2019}; from mobility surveys \cite{TAMBLAY201644}; through clustering techniques \cite{Cluster_1}; by means of neural networks or signal-processing methods \cite{NYU_OD}; and even from entropy-based approaches \cite{Lopez-Ospina_OD_entropy}. In cities where transportation systems record only entry points, strategies such as the use of exit tickets have been implemented to construct more accurate OD matrices \cite{Ferry_NY}. In the case of Bogotá’s public transportation system, previous studies have focused on a single transportation mode (e.g., BRT) and considered only two-step trips to construct the OD matrices \cite{Lotero_covid}.
\\[2mm]
Building on previous works and our results, we constructed an OD matrix whose elements $T_{ij}$ represent the number of users traveling from zone $i$ to zone $j$, considering all transportation modes integrated within the system. Moreover, all user trajectories throughout the day were included, not only those with two card validations, to generate the most complete OD matrix possible from the available data and, consequently, to achieve a deeper understanding of the system. Following the motif results presented in Fig. \ref{Fig_2}, we selected trajectories consisting of two to six validations, as supported by the information in Table \ref{Tabla_1} and the motif analysis. Using the same logic employed to identify mobility motifs, whereby each validation is treated as the destination of the preceding one, we first construct specific OD matrices for each trajectory length. These matrices, denoted OD-2 to OD-6, are generated from sequences of 2 to 6 successive access records. This approach yields a cleaner and more structured construction of the overall OD matrix. By combining these contributions, we obtain an integrated OD matrix that overcomes the limitation of restricting the analysis to two-step trajectories and provides a more comprehensive description of urban mobility in Bogotá. Although the OD-2 to OD-6 matrices are based on pairwise transitions, these transitions are systematically extracted from multi-step daily motifs rather than from isolated consecutive validations, allowing intermediate movements within the same day to be incorporated into the integrated representation.
\\[2mm]
The resulting OD-2 to OD-6 matrices were subsequently aggregated to obtain a final OD matrix representing the overall mobility of the city. This representation is shown in Fig. \ref{Fig_3}(a), which, with a size of $2{,}828 \times 2{,}828$ zones, captures the movement patterns of public transport users over the two-year study period. The color bar indicates the number of passengers that completed each trip. It is also worth noting that a large fraction of the OD matrix entries are zero: $55.2\%$ of all origin–destination pairs register no trips, indicating that users do not travel between many potential zone pairs. On average, a zone connects with 1264.8 other zones, although this value exhibits substantial variability, with a standard deviation of 742.4. For comparison, Fig. \ref{Fig_3}(b) presents the distance matrix of Bogotá, where the color bar represents the geographical distance between zones $i$ and $j$.
\\[2mm]

\subsection{Comparison of results using the Jensen--Shannon criterion}

Once the OD matrix has been constructed, it is necessary to establish a comparison criterion to verify that merging OD matrices from different trajectories is both valid and informative across semesters. For this purpose, we employ the Jensen–Shannon divergence, a statistical measure of similarity that generalizes the Kullback–Leibler divergence, defined for the probability distributions $P$ and $Q$ as \cite{KullbackLiebler_1951}
\begin{equation}\label{equ_KL}
	D_{\mathrm{KL}}(P \parallel Q) = \sum_{l} P_l \log \left( \frac{P_l}{Q_l} \right),
\end{equation}
which quantifies the dissimilarity between a probability distribution $P$ and a reference distribution $Q$, or equivalently, the information lost when $P$ is used to approximate $Q$. From Eq. (\ref{equ_KL}), the Jensen–Shannon divergence is defined as \cite{LinIEEE_1991}
\begin{equation}\label{equ_JS}
	D_{\mathrm{JS}}(P \parallel Q) = \frac{1}{2} D_{\mathrm{KL}}\!\left(P \parallel \frac{P+Q}{2}\right)
	+ \frac{1}{2} D_{\mathrm{KL}}\!\left(Q \parallel \frac{P+Q}{2}\right).
\end{equation}
This measure quantifies the similarity between two probability distributions, assigning values close to zero to equivalent distributions and values close to one to highly dissimilar ones. Since this criterion applies only to probability distributions, the OD matrix must first be transformed accordingly. In an OD matrix, rows correspond to origin zones and columns to destination zones, with each element $T_{ij}$ representing the number of trips from zone $i$ to zone $j$ during a given period. To convert it into a probability distribution, each element is normalized by the total number of recorded trips
\begin{equation}\label{equ_prob}
	p_{ij} = \frac{T_{ij}}{\sum_{i,j} T_{ij}},
\end{equation}
\begin{figure}[!t]
	\centering
	\includegraphics[width=1\linewidth]{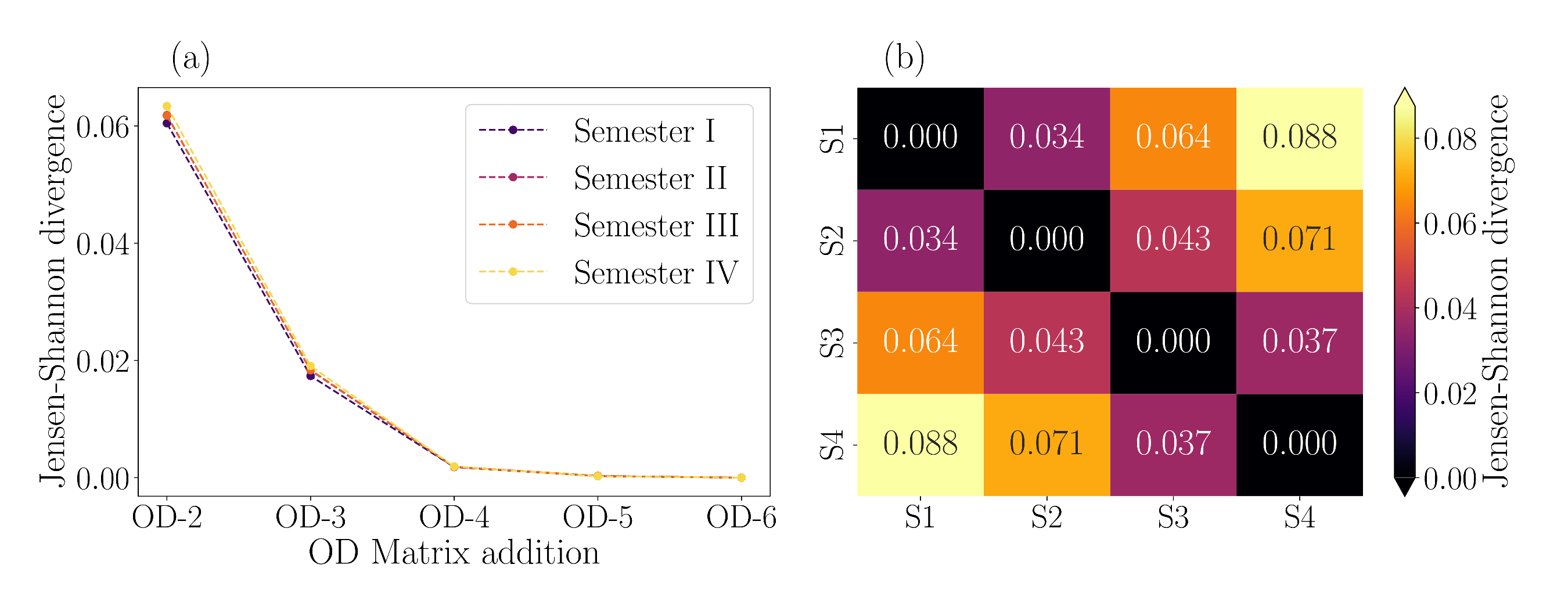}
	\caption{{\bf Comparative study of OD matrices using the Jensen--Shannon criterion.}  Based on the conversion of the OD matrix into probabilities $p_{ij}$ defined in Eq. (\ref{equ_prob}), the Jensen--Shannon divergence was used to compare OD matrices.  
		(a) Accumulated OD matrices: as additional trajectories are incorporated from OD-2 through OD-6, the matrix progressively captures more information, with OD-2 and OD-3 providing the largest contributions.  (b) Jensen--Shannon divergence across the four semesters $\mathrm{S1},\, \mathrm{S2},\,\mathrm{S3}$ and $\mathrm{S4}$ of the study, showing subtle temporal changes, with smaller differences observed between consecutive semesters.
	}
	\label{Fig_4}
\end{figure}
ensuring that $p_{ij} \geq 0$ and $\sum_{i,j} p_{ij} = 1$. After this conversion, the Jensen--Shannon divergence can be applied to quantify differences between OD matrices independently of the total number of trips.
\\[2mm]
Figure \ref{Fig_4} presents two complementary comparisons of the OD matrix based on the Jensen Shannon criterion. Figure \ref{Fig_4}(a) compares the complete OD matrix for each semester with matrices constructed from different numbers of accumulated trajectories. In all cases explored, the values of the Jensen Shannon divergence are small compared with its maximum value of 1. As additional trajectories are incorporated, the OD matrix becomes increasingly informative and the divergence approaches zero, confirming the validity of the proposed construction method. Notably, the matrices corresponding to two and three trajectories (OD-2 and OD-3) contribute most to the total OD matrix, as they account for the largest number of trips. Overall, these results indicate that the adopted procedure effectively enriches the mobility analysis.
\\[2mm]
On the other hand, Fig.~\ref{Fig_4}(b) shows the Jensen--Shannon divergence for OD matrices computed across different semesters, illustrating how public transport mobility in Bogotá evolved over time. For reference, we also evaluated a null model obtained by randomly shuffling the entries of each matrix, which produced values close to 0.9. Since fewer than 55.2\% of zones are mutually connected in the full two-year OD matrix, such perturbations significantly increase the Jensen--Shannon divergence, indicating that low values correspond to minimal changes in the structural organization of the matrix. These small variations may be associated with factors such as infrastructure works that required the closure of several BRT stations, the extension of the transfer time from 95 to 125 minutes, or the unification of fares across systems during the last three semesters. Nevertheless, despite these changes, the values remain very similar across semesters, indicating that mobility in the city retains a largely homogeneous character over time. 
\section{Results}
\subsection{Transition probabilities: modelling using L\'evy flights}
\label{Sec_prob_transicion}
Once reliable OD matrices are obtained, a detailed analysis of urban mobility can be performed. The mobility between city zones, described by OD matrices, can be represented as a spatial network. User movements across zones are then modeled as a dynamical process in which the transition probability $w_{i \to j}^{(\mathrm{OD})}$ from zone $i$ to zone $j$ is defined in terms of the OD matrix entries $T_{ij}$ as \cite{LoaizaPlosOne2019,Riascos_taxis}
\begin{equation}\label{wijOD}
	w_{i\to j}^{(\mathrm{OD})}= \frac{T_{ij}}{ k_i^{(\mathrm{out})}},
\end{equation}
where the out-degree $k_i^{(\mathrm{out})}=\sum_{\ell=1}^\mathcal{N}T_{i\ell}$ (with $\mathcal{N}$ denoting the number of zones) ensures the normalization $\sum_{\ell=1}^\mathcal{N} w_{i\to \ell}^{(\mathrm{OD})}=1$, guaranteeing that the total probability of traveling from zone $i$ to any other zone equals one.
\\[2mm]
To investigate the spatial dynamics between zones, we study the relation between transition probabilities $w_{i\to j}^{(\mathrm{OD})}$ and geographic distances $d_{ij}$. Several distance metrics can be used in this context; for instance, the Manhattan distance corresponds to the shortest path along the street network. The relation between users’ mobility intention, quantified by $w_{i\to j}^{(\mathrm{OD})}$, and distance $d_{ij}$ addresses an open question in the characterization of urban transport modes and remains little explored in OD-based analyses. In this study, transition probabilities are calculated from Eq. (\ref{wijOD}), while geographic distances are obtained from zone coordinates.
\\[2mm]
Moreover, different studies about human mobility suggest that the spatial dynamics of the system can be approximated by a Lévy flight model defined as $	w_{i\to j}^{\mathrm{(\mathrm{OD})}}\propto d_{ij}^{-\gamma}$ for $d_{ij}>0$ \cite{RiascosMateosPlos2017,LoaizaPlosOne2019,Riascos_taxis}. Lévy flights are a well-established framework for modeling mobility in continuous spaces, with applications ranging from human movement \cite{Brockmann2006,Gonzalez2008} and animal foraging \cite{BoyerPRS2006,ViswaBook,DaLuzPlos2017} to anomalous diffusion \cite{MetzlerPhysRep2000} and related processes \cite{RevModPhysZaburdaev2015}. In networks and other discrete settings, Lévy-flight dynamics were introduced in \cite{RiascosMateosLF2012} and further developed in several contexts \cite{FractionalBook2019,reviewjcn_2021}.
\\[2mm]
In this manner, the activity of users registered in the  OD matrix can be expressed in a simplified model capable of reproducing both short-range and long-range displacements, consistent with Lévy-flight behavior. A model with these characteristics was introduced by Riascos and Mateos to describe the movement of individuals among specific urban locations such as restaurants, universities, or public libraries \cite{RiascosMateosPlos2017}. In this framework, mobility is represented as random transitions among $\mathcal{N}$ locations, labeled $i = 1, 2, \ldots, \mathcal{N}$, which in our case correspond to the city zones used in the OD matrices. The transition probability $w_{i\to j}^{(\gamma)}$ for a hop from $i$ to $j$ is defined as \cite{RiascosMateosPlos2017}
\begin{equation}\label{wijmodel}
	w_{i\to j}^{(\gamma)}=\frac{\Omega_{ij}^{(\gamma)}}{\sum_{\ell=1}^{\mathcal{N}} \Omega_{i\ell}^{(\gamma)}},
\end{equation}
with
\begin{equation}\label{Omega_ij}
	\Omega_{ij}^{(\gamma)}=\left(d_0/d_{ij}\right)^\gamma \qquad \mathrm{for} \quad d_{ij}>0,
\end{equation}
where $\gamma$ is a positive real parameter and $d_0=1\,\mathrm{m}$ a reference distance.
\\[2mm]
To quantitatively compare the empirical transitions $\omega_{i\to j}^{(\mathrm{OD})}$ with those generated
by the distance-based model $w_{i\to j}^{(\gamma)}$, we interpret both matrices as probability
distributions over the set of ordered pairs $(i,j)$. This is achieved by flattening the matrices into the distributions
\begin{equation}
	P_{ij}(\gamma) = \frac{1}{\mathcal{N}} w_{i\to j}^{(\gamma)},
	\qquad
	Q_{ij}=\frac{1}{\mathcal{N}} w_{i\to j}^{\mathrm{(\mathrm{OD})}} .
	\label{Eq:Flattening}
\end{equation}
We then define a symmetrized measure $\mathcal{M}_{\mathrm{KLS}}(\gamma)$ based on the  Kullback--Leibler divergence
\begin{equation}
	\mathcal{M}_{\mathrm{KLS}}(\gamma)
	\equiv
	\frac{1}{2}
	\left[
	D_{\mathrm{KL}}\!\left(P(\gamma)\|Q\right)
	+
	D_{\mathrm{KL}}\!\left(Q\|P(\gamma)\right)
	\right],
	\label{Eq:KLS_gamma}
\end{equation}
where $D_{\mathrm{KL}}(A\|B)$ denotes the Kullback Leibler divergence defined in Eq. (\ref{equ_KL}), with the summation restricted to pairs $(i,j)$ for which both $A_{ij}$ and $B_{ij}$ are strictly positive, thereby ensuring numerical stability. The quantity $\mathcal{M}_{\mathrm{KLS}}(\gamma)$ provides a quantitative measure of the discrepancy between empirical and modeled transition probabilities, and its minimum identifies the exponent $\gamma^\star$ that best reproduces the observed mobility patterns.
\\[2mm]
\begin{figure}[t!]
	\centering
	\includegraphics[width=1\linewidth]{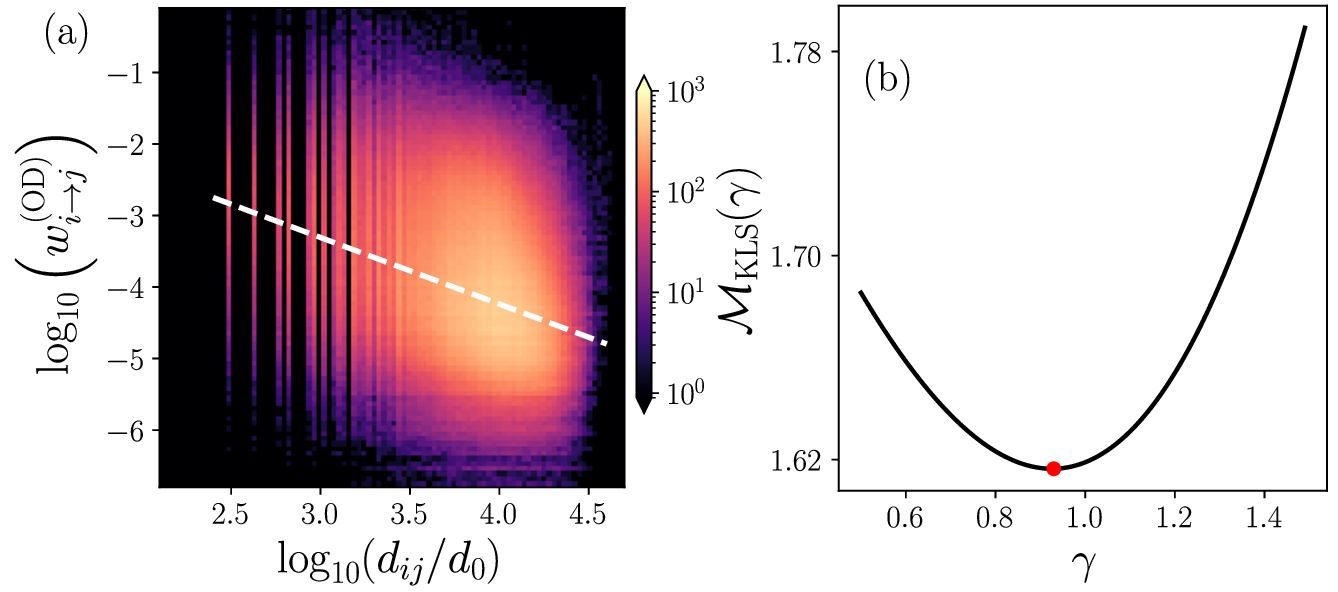}
	\caption{\label{Fig_5} 
		{\bf Relationship between the transition probability $\omega_{i \to j}^{(\mathrm{OD})}$ and the geographic distance $d_{ij}$ between zones $i$ and $j$.}  
		Analysis of SITP users in the Semester 2023-II. (a) Bidimensional histograms are constructed from $\log_{10}\omega_{i\to j}^{(\mathrm{OD})}$ and $\log_{10}(d_{ij}/d_0)$, with $d_0=1\,\mathrm{m}$ as a reference distance. Frequencies $f(d_{ij},\omega_{i\to j}^{(\mathrm{OD})})$  are encoded in the color bar, showing hexagonal bin counts on a logarithmic scale. The dashed line depicts the linear fit obtained from the distribution of pairs $\left(\log_{10}(d_{ij}/d_0), \log_{10}\omega_{i\to j}^{(\mathrm{OD})}\right)$. (b) Measure $\mathcal{M}_{\mathrm{KLS}}(\gamma)$ as a function of $\gamma$, defined in Eq. (\ref{Eq:KLS_gamma}), comparing the empirical transition probabilities $w_{i \to j}^{(\mathrm{OD})}$ with the model probabilities $w_{i \to j}^{(\gamma)}$. The minimum of $\mathcal{M}_{\mathrm{KLS}}(\gamma)$ represented with a dot identifies the optimal exponent $\gamma^\star = 0.93$.
	}
\end{figure}
In Fig. \ref{Fig_5} we examine the relationship between the information contained in the OD matrix and Lévy flight behavior for the dataset corresponding to Semester 2023-II. Figure \ref{Fig_5}(a) displays $\log_{10} w_{i \to j}^{(\mathrm{OD})}$ as a function of $\log_{10} d_{ij}$. To quantify the underlying trend, we construct a bidimensional histogram of the pairs $(x,y)$ defined by $\left(\log_{10}(d_{ij}/d_0), \log_{10} w_{i \to j}^{(\mathrm{OD})}\right)$, considering only nonzero values of $d_{ij}$ and $w_{i \to j}^{(\mathrm{OD})}$, with $i,j = 1,2,\ldots,\mathcal{N}$ and the reference distance set to $d_0 = 1\,\mathrm{m}$. The distribution of data points reveals that a linear relation provides a good approximation to the observed trend, leading to the fitting form
\begin{equation}\label{fit_log_log}
	\log_{10}\left(w_{i \to j}^{(\mathrm{OD})}\right)
	= C - \gamma \log_{10}\left(d_{ij}/d_0\right),
\end{equation}
from which we obtain $\gamma = 0.92$ for Semester 2023-II.
\\[2mm]
To independently determine the optimal exponent $\gamma^\star$, Fig. \ref{Fig_5}(b) shows the values of  $\mathcal{M}_{\mathrm{KLS}}(\gamma)$ defined in Eq. (\ref{Eq:KLS_gamma}) as a function of $\gamma$. This measure compares the transition probabilities inferred from the data, $w_{i \to j}^{(\mathrm{OD})}$, with those generated by the model, $w_{i \to j}^{(\gamma)}$. The results exhibit a well-defined minimum at $\gamma^\star = 0.93$. Thus, both the linear fit of the empirical distribution and the comparison between transition matrices yield consistent estimates for $\gamma$. In the following analysis, we adopt $\gamma^\star$, as it is obtained from a direct comparison between the empirical and modeled transition matrices. Finally, the results in Fig. \ref{Fig_5}(b) indicate that, unlike other transportation systems analyzed using different methodologies, the value $\gamma^\star \approx 0.93$ obtained for Bogotá lies at the lower
end of the range reported for urban mobility in other cities, where estimates
typically fall between $\gamma \approx 1.0$ and $2.2$ and often exhibit saturation
effects at large distances due to fare zoning or spatial constraints
\cite{RiascosMateosPlos2017,LoaizaPlosOne2019,Riascos_taxis}, highlighting the distinctive
role of long-range displacements in a highly connected single-fare transport
system.

\subsection{Monte Carlo simulation of the model}
\label{Sec_MonteCarlo}
\begin{figure}[t!]
	\centering
	\includegraphics[width=1\linewidth]{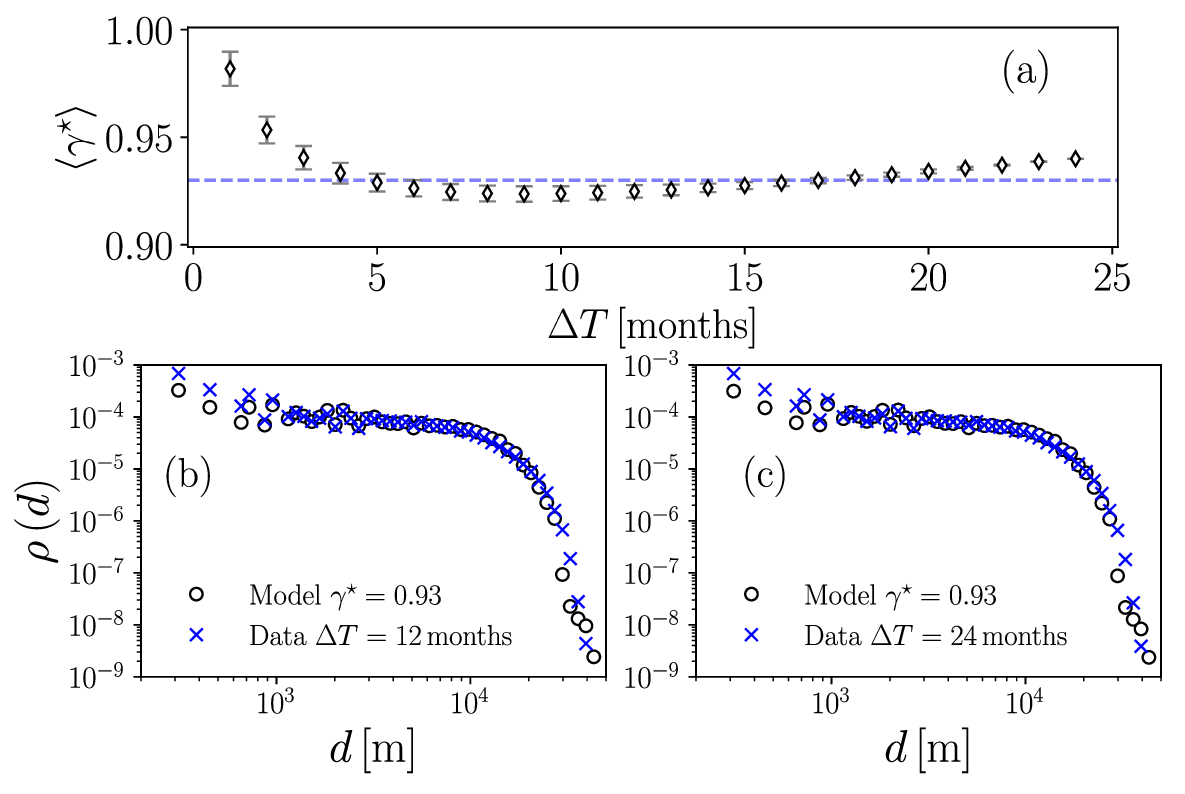}
	\caption{
		{\bf Statistical analysis of displacements between zones in the system.} (a) Average value $\langle \gamma^\star \rangle$ obtained by minimizing the measure $\mathcal{M}_{\mathrm{KLS}}(\gamma)$ for different temporal windows $\Delta T$, computed over all starting months $T$. Error bars indicate the corresponding standard deviation. The horizontal dashed line indicates the value $\gamma^\star = 0.93$, obtained by averaging over all temporal windows. 
		(b) Probability density $\rho(d)$ of the geographic distance $d$ between origin and destination zones obtained from mobility records and Monte Carlo simulations based on the model in Eq. (\ref{wijmodel}) using $\gamma^\star = 0.93$, for a temporal window $\Delta T = 12$ months corresponding to the first year of data.  (c) Same as panel (b), but for a temporal window $\Delta T = 24$ months.
	}
	\label{Fig_6} 
\end{figure}

The model proposed in Eq. (\ref{wijmodel}) captures the collective mobility dynamics through the single parameter $\gamma^\star$. In this section, we explore whether the inferred value of $\gamma^\star$ provides a coherent strategy for modeling transitions between zones and, consequently, for understanding the global spatial dynamics of the system. 
\\[2mm]
First, using the 24 months of available data, we construct OD matrices for different temporal windows $\Delta T$, measured in months, following the procedure described in Sec. \ref{Sec_ODmatrix}. For each $\Delta T$, we compute the empirical matrices with elements $w_{i \to j}^{(\mathrm{OD})}$ defined in Eq. (\ref{wijOD}) and the corresponding model matrices $w_{i \to j}^{(\gamma)}$ given in Eq. (\ref{wijmodel}). We then determine the optimal value $\gamma^\star$ by minimizing the measure $\mathcal{M}_{\mathrm{KLS}}(\gamma)$ defined in Eq. (\ref{Eq:KLS_gamma}). This analysis is repeated for different starting months $T = 1, 2, \ldots, 24$ and for all temporal windows $\Delta T = 1, 2, \ldots, 24$. The results are summarized in Fig. \ref{Fig_6}(a), which reports the average value $\langle \gamma^\star \rangle$, with error bars indicating the corresponding standard deviation. The results shown in Fig. \ref{Fig_6}(a) indicate that, for temporal windows $\Delta T < 5$ months, the values of $\langle \gamma^\star \rangle$ exhibit substantial variability, whereas they stabilize in the range $6 \leq \Delta T \leq 18$ months. Averaging over all temporal windows considered, we find that the data are well described by a single value $\gamma^\star = 0.93$, which is indicated in the figure with a dashed line.
\\[2mm]
We next test the model through Monte Carlo simulations. To this end, we generate trajectories for multiple users by selecting initial zones at random with probabilities proportional to their out degrees $k_m^{(\mathrm{out})}$, for $m = 1, 2, \ldots, \mathcal{N}$, which quantify the relative importance of each zone in the network. From each selected origin, the subsequent zone is drawn according to the transition probabilities defined in Eq. (\ref{wijmodel}), using the value $\gamma^\star = 0.93$ obtained from the analysis in Fig. \ref{Fig_6}(a). This procedure is iterated until the total number of simulated nonzero displacements equals that observed in the empirical OD matrices.
\\[2mm]
For each simulated transition, the geographic distance $d$ between the origin and destination zones is obtained from the distance matrix shown in Fig. \ref{Fig_3}(b). Figure \ref{Fig_6}(b) shows the probability density $\rho(d)$ of interzonal distances obtained from both empirical mobility records with $d > 0$ and Monte Carlo simulations based on the model in Eq. (\ref{wijmodel}), for a temporal window $\Delta T = 12$ months corresponding to the first year of data. The comparison shows that the empirical distribution $\rho(d)$ is well reproduced by transitions generated from $w_{i \to j}^{(\gamma)}$ with $\gamma^\star = 0.93$. In Fig. \ref{Fig_6}(c), we repeat the same analysis for $\Delta T = 24$ months, which considers the full dataset and uses the OD matrix shown in Fig. \ref{Fig_3}(a).
\\[2mm]
Our findings indicate that a temporal window of at least six months is sufficient to obtain a reliable estimate of the exponent characterizing the L\'evy flight dynamics from the OD matrix. Moreover, the similarity of the values of $\gamma^\star$ across different semesters reflects both the internal consistency of the analysis and the relative stability of mobility patterns in Bogotá over time. The largest deviations observed for temporal windows $\Delta T \geq 16$ months are likely associated with transient modifications of the transport system, primarily induced by the ongoing construction of the city’s metro infrastructure, which affects routes, stations, and transfer times in several high-demand areas. Nevertheless, at the global scale, the results indicate that urban mobility adapts in a robust manner despite these localized changes. This robustness persists despite broader structural reforms implemented up to 2023, including the elimination of low-demand routes, route consolidation, and improvements in route planning, as well as the unification of the fare structure in early 2024.
\subsection{Analysis of the dynamics on weekends}
\begin{figure}[t!]
	\centering
	\includegraphics[width=0.97\textwidth]{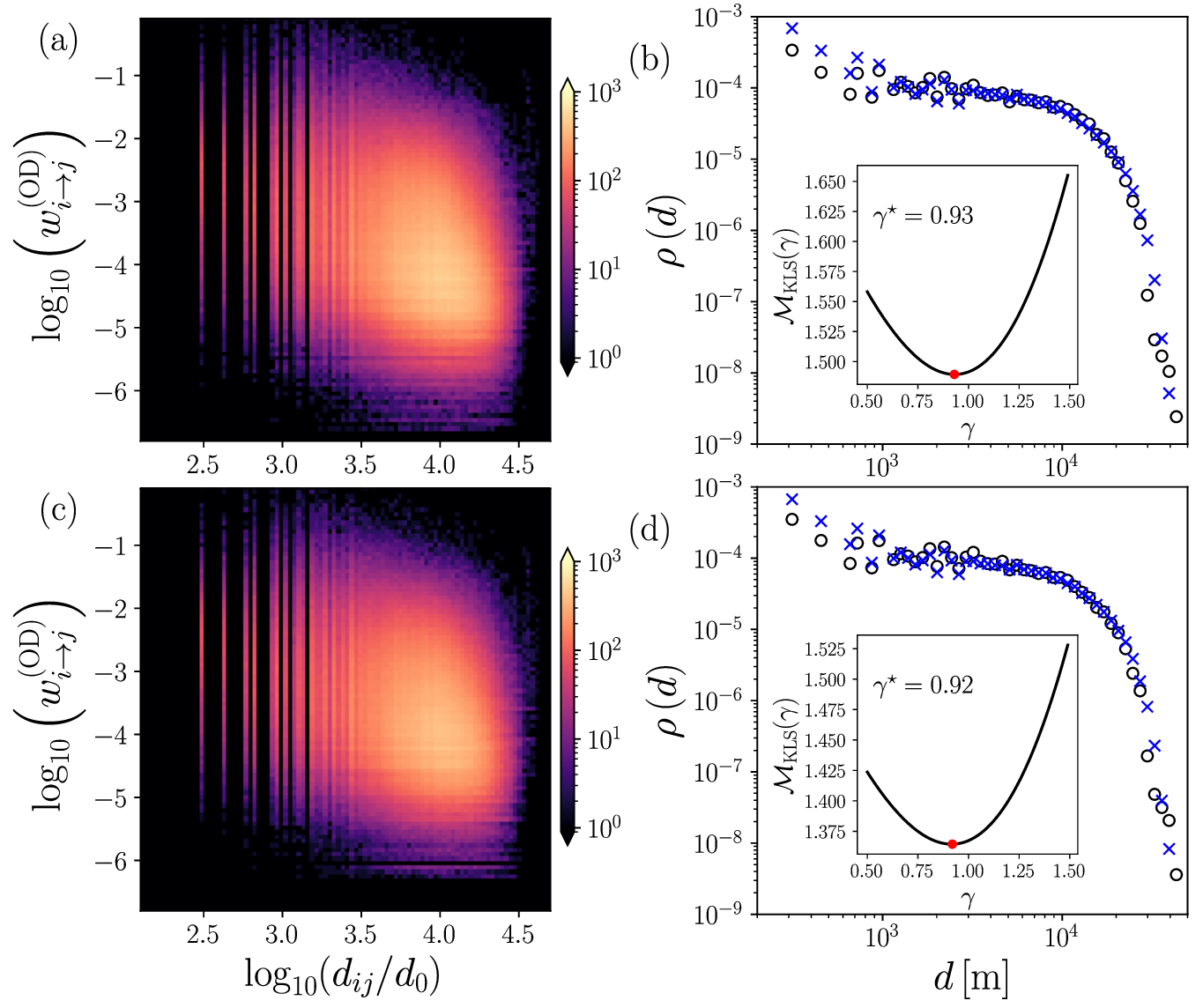}
	\caption{{\bf Weekend mobility dynamics in Bogotá.} (a) Bidimensional histogram of $\log_{10} w^{(\mathrm{OD})}_{i \to j}$ as a function of $\log_{10}(d_{ij}/d_0)$ for Saturdays, constructed using two years of data. (b) Probability density $\rho(d)$ of interzonal distances for Saturdays obtained from empirical mobility records and Monte Carlo simulations based on the Lévy flight model in Eq. (\ref{wijmodel}) using $\gamma^\star = 0.93$. The inset shows the minimization of the measure $\mathcal{M}_{\mathrm{KLS}}(\gamma)$ defined in Eq. (\ref{Eq:KLS_gamma}). (c) and (d) Same analysis as in panels (a) and (b), respectively, but for Sundays. In this case, the minimization of $\mathcal{M}_{\mathrm{KLS}}(\gamma)$ yields $\gamma^\star = 0.92$.}
	\label{Fig_7}
\end{figure}

The analysis presented so far has focused on weekdays, when public transport
usage is dominated by regular commuting patterns and the system experiences its
highest demand. Nevertheless, a natural question is whether the spatial dynamics
reported above remain valid during weekends, when travel purposes are more
diverse and daily routines are less structured. Addressing this point provides
an additional test of the robustness of the Lévy-flight description. To this end, we repeat the analysis of Secs. \ref{Sec_prob_transicion} and \ref{Sec_MonteCarlo} using weekend data, treating Saturdays and Sundays separately and considering the complete two-year dataset. As before, OD matrices are constructed using the motif-based approach, and transition probabilities are analyzed as a function of the
geographic distance between zones.
\\[2mm]
Figure~\ref{Fig_7}(a) shows the bidimensional histogram of the transition probabilities $w^{(\mathrm{OD})}_{i\to j}$ as a function of distance for Saturdays. As in the weekday case, the data exhibit a linear trend in log--log scale, indicating a power-law decay with distance. The corresponding comparison between empirical distance distributions and Monte Carlo simulations based on Eq. (\ref{wijmodel}) is shown in Fig. \ref{Fig_7}(b), the inset depicts the curve $\mathcal{M}_{\mathrm{KLS}}(\gamma)$ where the minimum $\gamma^\star \approx 0.93$ is obtained. The agreement between data and simulations remains.  Similarly,  Figs. \ref{Fig_7}(c) and (d) report the same analysis for Sundays. Despite a lower overall mobility demand and different temporal activity patterns, the spatial structure of displacements is again well captured by the Lévy-flight model with $\gamma^\star = 0.92$. These results indicate that, while trip timing and purposes vary substantially between weekdays and weekends, the underlying spatial dynamics governing interzonal movements remain remarkably stable.
\\[2mm]
Taken together, the results in Fig. \ref{Fig_7} for the weekend analysis supports the interpretation that the observed power-law decay of transition probabilities is not a byproduct of weekday commuting alone. Instead, it reflects persistent features of the city
and its transport network that shape mobility across different temporal regimes.
\\[2mm]
Finally, the results found in this section highlight distinctive features of mobility in Bogotá.  A key characteristic of the system is that users may travel between any two points with a single fare, which facilitates the occurrence of long-distance trips, whether by necessity or as a consequence of the network’s high level of connectivity. In contrast, short trips are relatively infrequent, as the fare remains comparatively high with respect to the minimum wage. In these cases, many users prefer to walk, since transportation costs represent a significant portion of their daily expenses. Moreover, unlike in other cities, fare differentials have not been implemented, not even for areas such as the airport, further reinforcing the incentive for long trips within the system.
\\[2mm]
A value of $\gamma^\star\approx 0.93$ places the system in the superdiffusive regime. This regime is characterized by a high probability of long-distance displacements, indicating a heterogeneous pattern of spatial exploration by users. In contrast with a standard Brownian process, understood here as a normal random walk with local hops to nearest neighbor zones, this behavior reveals a stronger tendency toward non-local movements and the absence of a dominant spatial scale. As a result, mobility does not concentrate around a characteristic radius; instead, the system displays patterns that connect both nearby and distant zones, thereby promoting broad spatial mixing. From an urban perspective, this suggests that the city’s morphology and the structure of its public transport network, particularly the availability of transfers, support users’ ability to traverse long distances across the territory.
\\[2mm]
From the standpoint of urban planning, the results presented in this section reveal spatial inequalities associated with the prevalence of long-range movements, which place a substantial load on the public transport system. Even so, the system continues to fulfill its essential role of ensuring inter-zonal connectivity, including for long-distance trips. Therefore, when designing or planning new transport modes, it is crucial to recognize that mobility patterns in Bogotá do not follow a local dynamics, but instead reflect power-law behavior characteristic of superdiffusive processes. This highlights the importance of explicitly incorporating long-distance displacements into the formulation of mobility policies and the development of urban transport strategies.

\section{Conclusions}

In this study, we analyzed the behavior of public transport users in Bogotá, identifying the main mobility patterns, particularly recurrent travel routines with their associated schedules and high-frequency zones. By employing motifs, we characterized the intermediate structure of daily trajectories, showing that users often perform sequences involving more than two observed entries. Although the entry-only nature of the data prevents us from identifying the final destination, and therefore a return to the initial zone cannot be ruled out, these motifs reveal that daily mobility is not restricted to a simple two-step pattern such as home $\to$ work $\to$ home, but typically includes multiple stops across different zones. Incorporating these multi-step patterns enables a more accurate reconstruction of user movements than methods based solely on successive pairs of validations, thereby improving the resolution of the OD matrix. This approach yielded an OD representation whose consistency was validated through the Jensen--Shannon divergence, demonstrating the relevance of the proposed methodology for analyzing urban mobility.
\\[2mm]
From the OD matrix constructed solely from entry records, we identify a clear relationship between transition probabilities and the geographical distance separating zones $i$ and $j$. This observation enables a modeling of user displacements in terms of Lévy flight dynamics, a description that is further supported by Monte Carlo simulations. The estimated values of $\gamma^\star$ remain stable across different temporal windows used to construct the OD matrices, indicating the robustness of the mobility patterns despite internal variations in the system. Taken together, these results provide a consistent and quantitative framework for studying urban mobility and the dynamics of public transportation in the city.
\\[2mm]
Our findings contribute to a deeper understanding of the complexity of Bogotá’s multimodal transport system by offering a more accurate representation of human movements based on motifs. Furthermore, the characterization of displacements through Lévy flights represents an innovative contribution in the local context, enabling a detailed description of user mobility dynamics. Consequently, this work provides valuable tools both for optimizing public transportation planning and for conducting comparative analyses with other mobility systems, whether similar or distinct.
\\[2mm]
Nevertheless, the study is limited by the methodology adopted for constructing OD matrices, which relies exclusively on user access validation records. This limitation opens opportunities for exploring alternative approaches to OD matrix generation. Finally, the results highlight the research potential of cities with less developed public transportation infrastructure, such as Bogotá, encouraging future studies aimed at deepening our understanding of these dynamics and contrasting them with those of major modern urban centers.
\section*{References}


\providecommand{\newblock}{}

\end{document}